\documentclass[useAMS,usenatbib]{mn2e}
\usepackage{graphicx}
\usepackage[T1]{fontenc}
\usepackage{aecompl}
\title[Basic properties of Fermi blazars and the ``blazar sequence'']{Basic properties of Fermi blazars and the ``blazar sequence''}
\author[Dingrong Xiong, Xiong Zhang, Jinming Bai and Haojing Zhang]{Dingrong Xiong$^{1,2}$, Xiong Zhang$^{3}\thanks{E-mail: ynzx@yeah.net}$, Jinming Bai$^{1}$ and Haojing Zhang$^{3}$\\
$^{1}$National Astronomical Observatories/Yunnan Observatories, Chinese Academy of Sciences, Kunming 650011, China\\
$^{2}$The Graduate School of Chinese Academy of Sciences, Beijing
100049, China\\
$^{3}$Department of Physics, Yunnan Normal University, Kunming
650500, China}
\begin{document}

\pagerange{\pageref{firstpage}--\pageref{lastpage}} \pubyear{2002}

\maketitle

\label{firstpage}

\begin{abstract}
By statistically analyzing a large sample which includes blazars of
Fermi detection (FBs) and non-Fermi detection (NFBs), we find that
there are significant differences between FBs and NFBs for redshift,
black hole mass, jet kinetic power from ``cavity'' power, broad-line
luminosity, and ratio of core luminosity to absolute V-band
magnitude ($R_{\rm v}$), but not for ratio of radio core to extended
flux ($R_{\rm c}$) and Eddington ratio. Compared with NFBs, FBs have
larger mean jet power, $R_{\rm c}$ and $R_{\rm v}$ while smaller
mean redshift, black hole mass, broad-line luminosity. These results
support that the beaming effect is main reason for differences
between FBs and NFBs, and that FBs are likely to have a more
powerful jet. For both Fermi and non-Fermi blazars, there are
significant correlations between jet power and the accretion rate
(traced by the broad-emission-lines luminosity), between jet power
and black hole mass; for Fermi blazars, the black hole mass does not
have significant influence on jet power while for non-Fermi blazars,
both accretion rate and black hole mass have contributions to the
jet power. Our results support the ``blazar sequence'' and show that
synchrotron peak frequency ($\nu_{\rm peak}$) is associated with
accretion rate but not with black hole mass.
\end{abstract}

\begin{keywords}
radiation mechanisms: nonthermal -- galaxies: active -- BL Lacertae
objects: general -- quasars: general -- gamma-rays: theory --
X-rays: general
\end{keywords}

\section{Introduction}

Blazars are the most extreme active galactic nuclei (AGN) pointing
their jets in the direction of the observer, and characterised by
extreme variability in their radio cores, high and variable
polarization, superluminal jet speeds and compact radio emission
(Angel \& Stockman 1980; Urry \& Padovani 1995). Relativistic
beaming of radiation is generally invoked to explain the extreme
properties (Madau, Ghisellini \& Persic 1987). Since the launch of
the Fermi satellite, we have entered in a new era of blazars
research (Abdo et al. 2009a, 2010a, 2012). Up to now, the Large Area
Telescope (LAT) has detected hundreds of blazars because it has
about 20 fold better sensitivity than its predecessor EGRET in the
0.1$-$100 GeV energy rang. According to the second catalogue of AGN
(2LAC, Ackermann et al. 2011), blazars are the brightest and the
most dominant population of AGN in the $\gamma$-ray sky. Many
answers have been proposed to explain the question: ``why are some
sources $\gamma$-ray loud and others $\gamma$-ray quiet?''
Generally, Doppler boosting, apparent speeds, very long baseline
interferometry (VLBI) core flux densities, brightness temperatures
and polarization are likely to be the important answers for this
question (Jorstad et al. 2001; Taylor et al. 2007; Kovalev et al.
2009; Lister et al. 2009a, 2009b; Savolainen et al. 2010; Piner et
al. 2012; Pushkarev et al. 2012; Linford et al. 2011, 2012; Wu et
al. 2014). Ghisellini et al. (2009b) studied general physical
properties of bright Fermi blazars. They modeled the spectral energy
distribution (SED) using a one zone leptonic model and confirmed the
relations of the physical parameters with source luminosity which
are at the origin of the blazar sequence. In these blazars they
argued that the jet must be proton dominated, and that the total jet
power is of the same order of (or slightly larger than) the disk
luminosity. In our earlier work (Xiong \& Zhang 2014; hereafter
XZ14), by compiling a large sample of clean blazars of 2LAC, we have
analyzed intrinsic $\gamma$-ray luminosity, black hole mass,
broad-line luminosity, jet kinetic power, and got that intrinsic
$\gamma$-ray luminosity with broad-line luminosity, black hole mass
and Eddington ratio have significant correlations; for almost all BL
Lacs, $P_{\rm jet}>L_{\rm disk}$ while for most of FSRQs, $P_{\rm
jet}<L_{\rm disk}$; the ``jet-dominance'' (parameterized as
$\frac{P_{\rm jet}}{L_{\rm disk}}$) is mainly controlled by the
bolometric luminosity. Recently, using the infrared colors of
Wide-Field Infrared Survey Explorer (WISE), Massaro et al. (2012)
have developed and successfully applied a new association method to
recognize $\gamma$-ray blazar candidates. However, at present, due
to Doppler boosting effect or limit of small sample, it still is
unclear whether the differences between $\gamma$-ray loud and
$\gamma$-ray quiet blazars are related with intrinsic properties.

Blazars are often divided into two subclasses of BL Lacertae objects
(BL Lacs) and flat spectrum radio quasars (FSRQs). FSRQs have strong
emission lines, while BL Lacs have only very weak or non-existent
emission lines. The classic division between FSRQs and BL Lacs is
mainly based on the equivalent width (EW) of the emission lines.
Objects with rest frame EW$>5$ {\AA} are classified as FSRQs (e.g.
Scarpa \& Falomo 1997; Urry \& Padovani 1995). Many authors have
proposed that EW alone is not a good indicator of the distinction
between the two classes of blazars (Scarpa \& Falomo 1997;
Ghisellini et al. 2011; Sbarrato et al. 2012, 2014; Giommi et al.
2012, 2013; XZ14). Ghisellini et al. (2011) introduced a physical
distinction between the two classes of blazars, based on the
luminosity of the broad line region measured in Eddington units. The
dividing line is of the order of $L_{\rm{BLR}}/L_{\rm{Edd}}\sim
5\times10^{-4}$. The result also was confirmed by Sbarrato et al.
(2012) and XZ14. Giommi et al. (2012, 2013) suggested that blazars
should be divided in high and low ionization sources.

Fossati et al. (1998) and Ghisellini et al. (1998) originally
presented a unifying view of the SED of blazars and the blazar
sequence: a strong anti-correlation between bolometric luminosity
and synchrotron peak frequencies. The scenario has been the subject
of intense discussions (Giommi, Menna \& Padovani 1999;
Georganopoulos et al. 2001; Cavaliere \& D'Elia 2002; Padovani et
al. 2003; Maraschi \& Tavecchio 2003; Nieppola et al. 2006, 2008;
Xie et al. 2007; Ghisellini \& Tavecchio 2008; Ghisellini et al.
2009a, 2010; Meyer et al. 2011; Chen \& Bai 2011; Giommi et al.
2012; Finke et al. 2013). Ghisellini et al. (1998) interpreted the
spectral sequence as that a stronger radiative cooling suffered by
the emitting electrons of blazar of larger radiative energy density
causes a particle energy distribution with a break at lower
energies. A more theoretical blazar sequence is related $\gamma_{\rm
peak}$ (Lorentz factor of the peak of the electron distribution
which is responsible for the majority of emission at the two peaks
of the SED) to the amount of radiative cooling. Ghisellini \&
Tavecchio (2008) revisited the blazar sequence and proposed that the
power of the jet and SED of its emission are linked to the mass of
black hole and the accretion rate. Padovani (2007) pointed out that
three main tests about blazar sequence should be got through:
anti-correlation between bolometric luminosity and synchrotron peak
frequencies; non-existence of high peak frequencies of powerful
objects; high peak frequencies of BL Lacs should be more numerous
than low peak frequencies of blazars. Nieppola et al. (2008)
proposed that the blazar sequence disappears when the intrinsic
Doppler-corrected values are used. Giommi et al. (2012) showed that
the blazar sequence is a selection effect arising from the
comparison of shallow radio and X-ray surveys, and that high
synchrotron peak frequency- high radio power objects have never been
considered because their redshift is not measurable. Padovani,
Giommi \& Rau (2012) have studied the quasi-simultaneous near-IR,
optical, UV, and X-ray photometry of eleven $\gamma$-ray selected
blazars, and found four high power - high synchrotron peak blazars.
Ghisellini et al. (2009a, 2010), Abdo et al. (2010b) and Sambruna et
al. (2010) have got that the correlation between $\gamma$-ray
luminosity and photon index supports the blazar sequence. Chen \&
Bai (2011) confirmed that low power - low synchrotron peak blazars
have relatively lower black hole masses. Meyer et al. (2011)
revisited the blazar sequence and proposed the blazar envelope: FR
Is and most BL Lacs belong to weak jet population while low
synchrotron peaking blazars and FR IIs are strong jet population.
The Compton dominance, the ratio of the peak of the Compton to the
synchrotron peak luminosities, is essentially a redshift-independent
quantity and thus crucial to answer the blazar sequence. Finke
(2013) studied a sample of blazars from 2LAC and found that a
correlation exists between Compton dominance and the peak frequency
of the synchrotron component for all blazars, including ones with
unknown redshift.

In this paper, we constructed a large sample of blazars, including
Fermi blazars from XZ14 and non-Fermi blazars, and studied the
properties of Fermi blazars and the blazar sequence. The paper is
structured as follows: in Sect. 2, we present the samples; the
results are presented in Sect. 3; discussions and conclusions are
presented in Sect. 4. The cosmological parameters $H_{\rm 0}=70~
{\rm km~s^{-1}~Mpc^{-1}}$, $\Omega_{\rm m}=0.3$ and $\Omega_{\rm
\Lambda}=0.7$ have been adopted in this work. The energy spectral
index $\alpha$ is defined such that $F_{\rm
\nu}\propto\nu^{-\alpha}$.

\section{The samples}

The selection criteria for the sample were that we tried to select
the largest group of blazars included in BZCAT (Massaro et al. 2009:
the Roma BZCAT) with reliable broad line luminosity (used as a proxy
for disk luminosity), redshift, black hole mass and jet kinetic
power. The sample of Fermi blazars was directly from XZ14. Due to be
classed into non-clean 2LAC, the four fermi blazars (2FGL
J0204.0+3045; 2FGL J0656.2-0320; 2FGL J1830.1+0617; 2FGL
J2356.3+0432) in XZ14 were not included in our sample. In order to
have reliable sample, we did not consider the candidate blazars of
unknown type (BZU called in BZCAT) in our sample. Also for the same
reason, non-Fermi blazars, which were detected by EGRET or recorded
in 1LAC but missed in 2LAC, were not included in our sample. In
addition, to reduce the uncertainty, we tried to select the data
from a same paper and/or a uniform method. The detailed information
and calculating methods for broad line luminosity, black hole mass
are seen in XZ14. Cavagnolo et al. (2010) searched for X-ray
cavities in different systems including giant elliptical galaxies
and cD galaxies and estimated the jet power required to inflate
these cavities or bubbles, obtaining a tight correlation between the
``cavity'' power and the radio luminosity at 200-400 MHz
\begin{equation}
P_{\rm cav}\approx5.8\times10^{43}(\frac{P_{\rm radio}}{10^{40}{\rm
erg~s^{-1}}})^{0.7} {\rm erg~s^{-1}},
\end{equation}
which is continuous over $\sim6-8$ decades in $P_{\rm jet}$ and
$P_{\rm radio}$ and $P_{\rm jet}=P_{\rm cav}$. Making use of the
correlation between $P_{\rm jet}$ and $P_{\rm radio}$ from Cavagnolo
et al. (2010), Meyer et al. (2011) chose the low-frequency extended
luminosity at 300 MHz as an estimator of the jet power for blazars.
Their 300 MHz extended luminosity was extrapolated from 1.4 GHz
extended radio emission or obtained from spectral decomposition.
Following Meyer et al. (2011), Nemmen et al. (2012) estimated the
jet kinetic power for a large sample of Fermi blazars. We also used
Equation (1) to get jet kinetic power from ``cavity'' power (this is
what we mean when we refer to the jet kinetic power in the rest of
the paper). In our sample, in order to reduce uncertainty, the jet
kinetic power only is gained from extended 1.4 GHz radio data but
not from spectral decomposition. The extended 1.4 GHz radio data can
not be obtained for all blazars. The proportion of blazars of jet
kinetic power estimated from extended radio luminosity in all
sources was 28\%. From Wang et al. (2004), the uncertainty in the
$M_{\rm BH}-\sigma$ relation was $\leq0.21$ dex; the uncertainty on
the zero point of the line width-luminosity-mass relation was
approximately 0.5 dex; the $M_{\rm BH}-M_{\rm R}$ correlation for
quasar host galaxies had an uncertainty of 0.6 dex. So given the
intrinsic uncertainty of the different black hole mass estimators
and the heterogeneity of sample, we estimated that the individual BH
masses may have an uncertainty as large as $\sim$1 dex. The
uncertainty in $P_{\rm jet}$ was dominated by the scatter in the
correlation of Cavagnolo et al. (2010) and corresponded to 0.7 dex.
The uncertainties on broad line luminosity were based on the
standard deviation of Mg II/Ly${\rm \alpha}$, H${\rm \beta}$/Ly${\rm
\alpha}$ and C IV/Ly${\rm \alpha}$ in the composite quasar spectrum
of Francis et al. (1991) (Wang et al. 2004). We assumed that the
uncertainties on broad line luminosity were to be 0.5 dex for all
sources.

For the sake of exploring the blazar sequence, we also collected
and/or calculated the synchrotron peak frequency $\nu_{\rm peak}$
and the peak luminosity of the synchrotron component $L_{\rm peak}$.
The $\nu_{\rm peak}$ and $L_{\rm peak}$ of our Fermi blazars were
collected from Finke (2013) and Meyer et al. (2011), and the
$\nu_{\rm peak}$ and $L_{\rm peak}$ of non-Fermi blazar from
Nieppola et al. (2006, 2008), Meyer et al. (2011), Wu et al. (2009),
Aatrokoski et al. (2011). Generally, the SED was fitted by using a
simple third-degree polynomial function. However, many blazars were
lack of observed SED. Abdo et al. (2010c) have conducted a detailed
investigation of the broadband spectral properties of the
$\gamma$-ray selected blazars of the Fermi LAT Bright AGN Sample
(LBAS). They assembled high-quality and quasi-simultaneous SED for
48 LBAS blazars, and their results have been used to derive
empirical relationships that estimate the position of the two peaks
from the broadband colors (i.e. the radio to optical, $\alpha_{\rm
ro}$, and optical to X-ray, $\alpha_{\rm ox}$, spectral slopes) and
from the $\gamma$-ray spectral index. Ackermann et al. (2011) used
the empirical relationships for finding the peak frequency of
synchrotron component of 2LAC clean blazars from the slopes between
the 5 GHz and 5500 {\AA} flux, and between the 5500 {\AA} and 1 KeV
flux. Finke (2013) used their results for finding $\nu_{\rm peak}$.
The rest of authors fitted SED to obtain $\nu_{\rm peak}$. When
blazars were missed in the above literatures, we used the empirical
relationships of Abdo et al. (2010c) to find $\nu_{\rm peak}$.
Firstly, we collected the fluxes of the blazars at 5 GHz, 5500 {\AA}
and 1 KeV from BZCAT, NASA/IPAC Extragalactic Database: NED,
Veron-Cetty \& Veron (2010) and Ackermann et al. (2011). When more
than one flux or magnitude was found, we took the most recent one.
Apparent magnitude of optical V band can be converted into flux as
$\log S=\log S_{\rm 0}-0.4m_{\rm V}$ with flux $S_{\rm 0}=3.64$ KJy
and $S$ in units of KJy (Mead et al. 1990). All flux densities were
K-corrected according to $S_{\rm \nu}=S^{\rm {obs}}_{\rm
\nu}(1+z)^{\rm \alpha-1}$, where $\alpha$ was the spectral index and
$\alpha_{\rm R}=0.0$, $\alpha_{\rm O}=1.0$, and $\alpha_{\rm
X}=1.47$ for BL Lacs, and $\alpha_{\rm X}=0.87$ for FSRQs (Cheng et
al. 2000). The luminosity was calculated from the relation $L_{\rm
\nu}=4\pi{d_{\rm L}}^2S_{\rm \nu}$, and $d_{\rm L}$ was the
luminosity distance. The 1 KeV flux density was transferred from the
0.1-2.4 KeV flux density given in the BZCAT or NED using $F_{\rm
\nu}\propto\nu^{-\alpha}$. After then we used the Equation (1) of
Abdo et al. (2010c) to get $\alpha_{\rm ro}$ and $\alpha_{\rm ox}$.
Finally, we used the empirical relationships of Abdo et al. (2010c)
for finding the peak frequency of synchrotron component. For blazars
without X-ray flux, we adopted $\alpha_{\rm ro}$ to estimate
$\nu_{\rm peak}$ ($\log \nu_{\rm
peak}=17.5(\pm0.22)-6.29(\pm0.38)\alpha_{\rm ro}$ from linear
regression analysis: $r=-0.71, N=266, P<0.0001$). The proportion of
blazars (that the $\nu_{\rm peak}$ is estimated by the linear
regression equation) in all sources is 20\%. When the peak
luminosity of synchrotron component was not got from above
literatures, we used radio luminosity at 5 GHz to estimate the
synchrotron peak luminosity ($\log L_{\rm peak}=0.61(\pm0.02)\log
L_{\rm 5GHz}+19.67(\pm0.82)$ from linear regression analysis:
$N=245, r=0.9, P<0.0001$). The proportion of blazars (that the
$L_{\rm peak}$ is estimated by the linear regression equation) in
all sources is 60\%.

The ratio of the beamed radio core flux density to the unbeamed
extended radio flux density ($R_{\rm c}=\frac{S_{\rm core}}{S_{\rm
ext}}(1+z)^{\alpha_{\rm core}-\alpha_{\rm ext}}$ with $\alpha_{\rm
core}=0$, $\alpha_{\rm ext}=1$) has routinely been used as a
statistical indicator of Doppler beaming and thereby orientation
(Orr \& Browne 1982; Urry \& Padovani 1995; Kharb et al. 2010). In
order to compare Fermi blazars with non-Fermi blazars, we also
estimated $R_{\rm c}$ at 1.4 GHz. When more than one $R_{\rm c}$ was
obtained, we took the most recent one.

The relevant data for Fermi blazars can be seen in Table 1 of XZ14.
The relevant data for non-Fermi blazars were listed in Table 1 with
the following headings: column (1), name of the Roma BZCAT catalog;
column (2), other name; column (3) is right ascension (the first
entry) and declination (the second entry); column (4), redshift from
NED; column (5), logarithm of the synchrotron peak frequency (the
first entry) and logarithm of the peak luminosity of the synchrotron
component in units of ${\rm{erg~s^{-1}}}$ (the second entry); column
(6), logarithm of jet kinetic power in units of ${\rm{erg~s^{-1}}}$;
column (7), logarithm of black hole mass in units of ${\rm M_{\rm
\odot}}$ and references; column (8), logarithm of broad-line
luminosity in units of ${\rm{erg~s^{-1}}}$ and references; column
(9), logarithm of the ratio of the beamed radio core flux density to
the unbeamed extended radio flux density and references. For black
hole mass or broad-line luminosity, when more than one value was
obtained, we took the mean value.

In total, we have a sample containing 244 clean Fermi blazars (187
FSRQs and 57 BL Lacs) and 469 non-Fermi blazars (370 FSRQs and 99 BL
Lacs).

\section{The results}
\subsection{The distributions}

The redshift distributions of the various classes are shown in Fig.
1. From Fig. 2 of Roma-BZCAT, the redshift distributions of BL Lacs
are much closer than that of FSRQs and their distribution peaks at
$z\cong0.3$, whereas FSRQs show a broad maximum between 0.6 and 1.5.
There are only very few BL Lacs at redshift higher than 0.8. So the
redshift distributions from our sample are consistent with the
results of Roma-BZCAT. The redshift distributions for Fermi blazars
are $0<z<3.1$ and mean value is $1.008\pm0.04$; for non-Fermi
blazars, the redshift distributions are $0<z<3.95$ and mean value is
$1.13\pm0.04$. The mean values for Fermi BL Lacs, Fermi FSRQs,
non-Fermi BL Lacs and non-Fermi FSRQs are $0.45\pm0.05$,
$1.18\pm0.05$, $0.3\pm0.01$ and $1.35\pm0.04$ respectively. Through
nonparametric Kolmogorov-Smirnov (KS) test, we get that the redshift
distributions between all Fermi blazars and all non-Fermi blazars,
between Fermi BL Lacs and non-Fermi BL Lacs, between Fermi FSRQs and
non-Fermi FSRQs are significant difference (chance probability
$P=0.006, P=0.001, P=0.002$). Based on above results, it is shown
that compared with non-Fermi BL Lacs, Fermi BL Lacs have larger mean
redshift while compared with non-Fermi FSRQs, Fermi FSRQs have
smaller mean redshift.

\begin{figure}
\includegraphics[width=95mm, height=80mm]{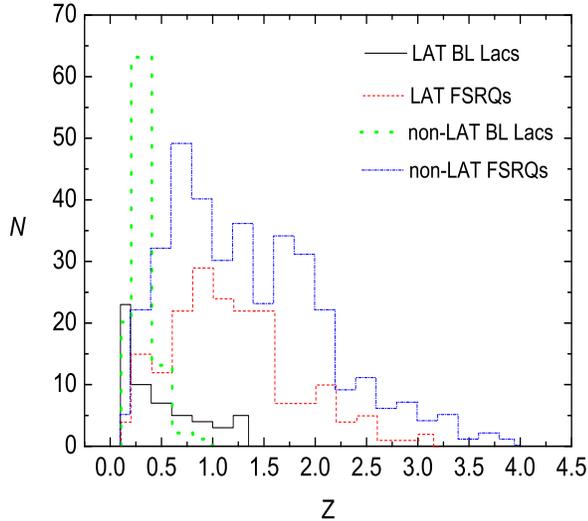}
\caption{Redshift distributions for BL Lacs detected by Fermi LAT
(LAT BL Lacs, black continuous line), FSRQs detected by Fermi LAT
(LAT FSRQs, red dashed line), BL Lacs of non $\gamma$-ray loud
(non-LAT BL Lacs, green dotted line), FSRQs of non $\gamma$-ray loud
(non-LAT FSRQs, blue dot--dashed line).} \label{figure 1}
\end{figure}

\begin{figure}
\includegraphics[width=95mm, height=80mm]{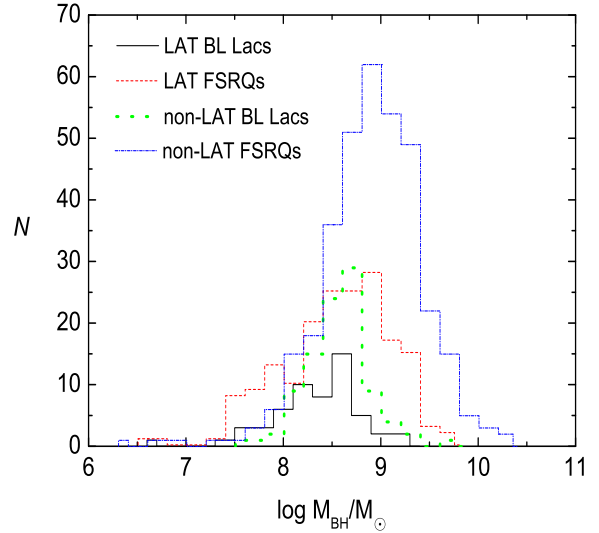}
\caption{Black hole mass distributions for LAT BL Lacs, LAT FSRQs,
non-LAT BL Lacs and non-LAT FSRQs. The meanings of different lines
are as same as Fig. 1. The black hole mass distributions for Fermi
blazars are $10^{6.5}-10^{9.8}$ ${\rm M_{\rm \odot}}$ and mean value
is $10^{8.5\pm0.03}$ ${\rm M_{\rm \odot}}$; for non-Fermi blazars,
the black hole mass distributions are $10^{6.35}-10^{10.24}$ ${\rm
M_{\rm \odot}}$ and mean value is $10^{8.82\pm0.02}$ ${\rm M_{\rm
\odot}}$. The mean values for Fermi BL Lacs, Fermi FSRQs, non-Fermi
BL Lacs and non-Fermi FSRQs are $10^{8.34\pm0.06}$ ${\rm M_{\rm
\odot}}$, $10^{8.55\pm0.04}$ ${\rm M_{\rm \odot}}$,
$10^{8.54\pm0.03}$ ${\rm M_{\rm \odot}}$ and $10^{8.9\pm0.03}$ ${\rm
M_{\rm \odot}}$ respectively.} \label{figure 1}
\end{figure}

\begin{figure}
\includegraphics[width=95mm, height=80mm]{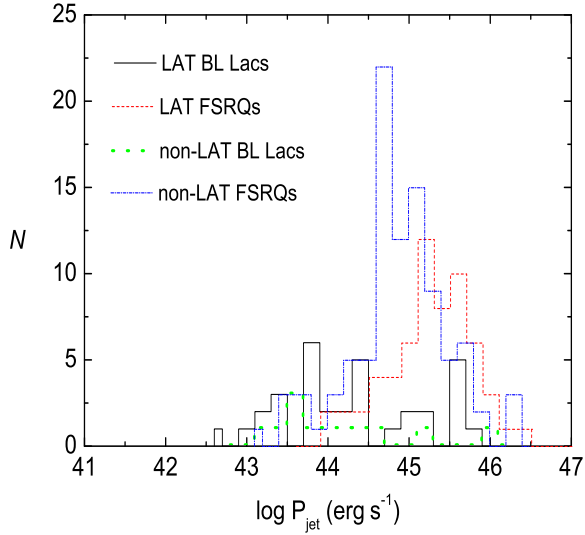}
\caption{Jet kinetic power distributions for LAT BL Lacs, LAT FSRQs,
non-LAT BL Lacs and non-LAT FSRQs. The meanings of different lines
are as same as Fig. 1. The jet kinetic power distributions for Fermi
blazars are $10^{42.6}-10^{46.5}\rm{erg~s^{-1}}$ and mean value is
$10^{44.92\pm0.08} \rm{erg~s^{-1}}$; for non-Fermi blazars, the jet
kinetic power distributions are $10^{43}-10^{46.4}\rm{erg~s^{-1}}$
and mean value is $10^{44.80\pm0.07} \rm{erg~s^{-1}}$. The mean
values for Fermi BL Lacs, Fermi FSRQs, non-Fermi BL Lacs and
non-Fermi FSRQs are $10^{44.32\pm0.15} \rm{erg~s^{-1}}$,
$10^{45.25\pm0.07} \rm{erg~s^{-1}}$, $10^{44.16\pm0.23}
\rm{erg~s^{-1}}$ and $10^{44.88\pm0.06} \rm{erg~s^{-1}}$
respectively.} \label{figure 1}
\end{figure}

\begin{figure}
\includegraphics[width=95mm, height=80mm]{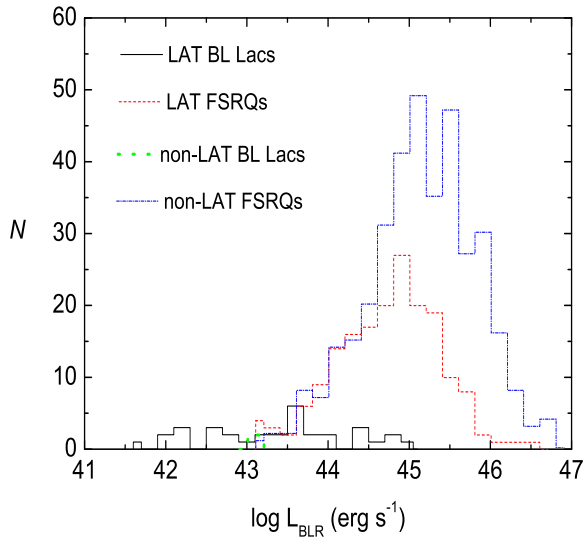}
\caption{Broad-line luminosity distributions for LAT BL Lacs, LAT
FSRQs, non-LAT BL Lacs and non-LAT FSRQs. The meanings of different
lines are as same as Fig. 1.} \label{figure 7}
\end{figure}

\begin{figure}
\includegraphics[width=95mm, height=80mm]{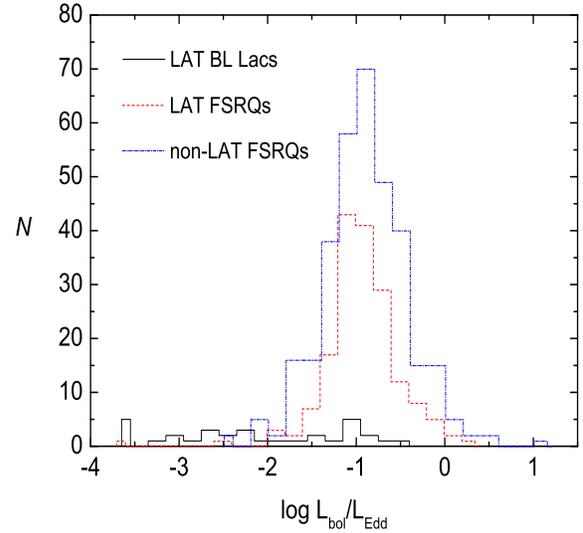}
\caption{Eddington ratio distributions for LAT BL Lacs, LAT FSRQs
and non-LAT FSRQs. The meanings of different lines are as same as
Fig. 1.} \label{figure 1}
\end{figure}

\begin{figure}
\includegraphics[width=95mm, height=80mm]{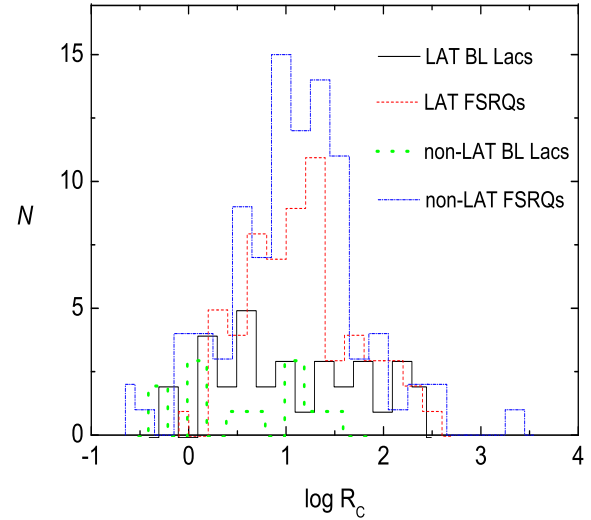}
\caption{Ratio of the beamed radio core flux density to the unbeamed
extended radio flux density distributions for LAT BL Lacs, LAT
FSRQs, non-LAT BL Lacs and non-LAT FSRQs. The meanings of different
lines are as same as Fig. 1.} \label{figure 10}
\end{figure}

\begin{figure}
\includegraphics[width=95mm, height=80mm]{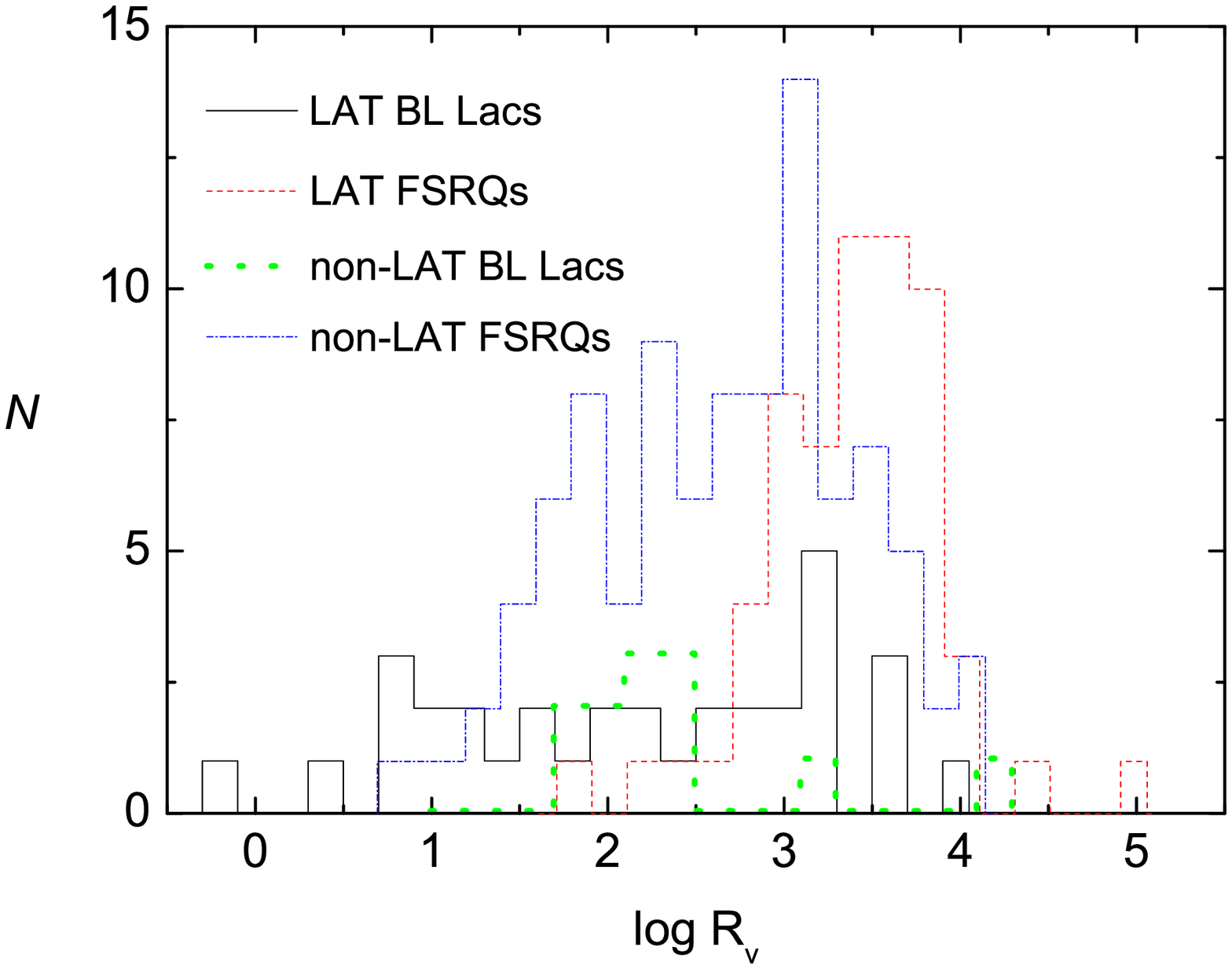}
\caption{The ratio of the radio core luminosity to the k-corrected
absolute V-band magnitude for LAT BL Lacs, LAT FSRQs, non-LAT BL
Lacs and non-LAT FSRQs. The meanings of different lines are as same
as Fig. 1.} \label{figure 10}
\end{figure}

The black hole mass distributions of the various classes are shown
in Fig. 2. The black hole mass distributions between all Fermi
blazars and all non-Fermi blazars, between Fermi BL Lacs and
non-Fermi BL Lacs, between Fermi FSRQs and non-Fermi FSRQs are
significant difference ($P<0.0001, P=0.011, P<0.0001$). So we can
get that compared with non-Fermi blazars, Fermi blazars have smaller
mean black hole mass.

The jet kinetic power distributions of the various classes are shown
in Fig. 3. The jet kinetic power distributions between all Fermi
blazars and all non-Fermi blazars, between Fermi FSRQs and non-Fermi
FSRQs are significant difference ($P=0.01, P=0.0002$). So Fermi
blazars have larger mean jet kinetic power than non-Fermi blazars.

The broad-line luminosity distributions of the various classes are
shown in Fig. 4. Because non-Fermi BL Lacs almost have not
broad-line data, we only compare broad-line luminosity distributions
between Fermi FSRQs and non-Fermi FSRQs. The mean values for Fermi
FSRQs and non-Fermi FSRQs are $10^{44.72\pm0.05} \rm{erg~s^{-1}}$
and $10^{45.14\pm0.03} \rm{erg~s^{-1}}$ respectively. From KS test,
the broad-line luminosity distributions between Fermi FSRQs and
non-Fermi FSRQs are significant difference ($P<0.0001$). We also
compare the Eddington ratio distributions between Fermi FSRQs and
non-Fermi FSRQs (see Fig. 5; $L_{\rm Edd}=1.3\times10^{38}(\frac{\rm
M}{\rm M_\odot}){\rm erg~s^{-1}}, L_{\rm bol}\approx10L_{\rm BLR}$
from Netzer (1990)). However, the result of KS test shows that they
do not have significant difference ($P=0.398$). The mean values for
Fermi FSRQs and non-Fermi FSRQs are $10^{-0.93\pm0.03}$ and
$10^{-0.91\pm0.03}$ respectively.

The distributions for ratio of the beamed radio core flux density to
the unbeamed extended radio flux density (core prominence parameter
$R_{\rm c}$) are shown in Fig. 6. The $R_{\rm c}$ distributions for
Fermi blazars are $10^{-0.3}<R_{\rm c}<10^{2.5}$ and mean value is
$10^{1.13\pm0.07}$; for non-Fermi blazars, the $R_{\rm c}$
distributions are $10^{-0.7}<R_{\rm c}<10^{3.5}$ and mean value is
$10^{1.02\pm0.07}$. The mean values for Fermi BL Lacs, Fermi FSRQs,
non-Fermi BL Lacs and non-Fermi FSRQs are $10^{1.09\pm0.13}$,
$10^{1.16\pm0.07}$, $10^{0.6\pm0.18}$ and $10^{1.08\pm0.07}$
respectively. The $R_c$ distributions between all Fermi blazars and
all non-Fermi blazars, between Fermi BL Lacs and non-Fermi BL Lacs,
between Fermi FSRQs and non-Fermi FSRQs are not significant
difference ($P=0.4, P=0.26, P=0.85$). Kharb et al. (2010) found that
the ratio of the radio core luminosity to the k-corrected optical
luminosity ($R_{\rm v}$) appears to be a better indicator of
orientation than the traditionally used radio core prominence
parameter ($R_{\rm c}$). Wills \& Brotherton (1995) defined $R_{\rm
v}$ as the ratio of the radio core luminosity to the k-corrected
absolute V-band magnitude ($M_{\rm abs}$): $\log R_{\rm v}=\log
(L_{\rm core}/L_{\rm opt})=(\log L_{\rm core}+M_{\rm
abs}/2.5)-13.7$, where $M_{\rm abs}=M_{\rm v}-k$, and the
k-correction is, $k=-2.5\log(1+z)^{1-\alpha_{\rm opt}}$ with the
optical spectral index, $\alpha_{\rm opt}=0.5$. Making use of the
above Equations, we obtain $R_{\rm v}$ for our sample. The
distributions for $R_{\rm v}$ are shown in Fig. 7. The mean values
of $R_{\rm v}$ for Fermi blazars and non-Fermi blazars are
$10^{2.95\pm0.1}$ and $10^{2.63\pm0.07}$ respectively. The $R_{\rm
v}$ distribution between all Fermi blazars and all non-Fermi blazars
is significant difference ($P=0.001$). Therefore, Fermi blazars have
larger mean $R_{\rm v}$ than non-Fermi blazars, which supports that
compared with non-Fermi blazars, Fermi blazars are more beamed.

\subsection{Black hole mass vs jet power}

Figure 8 shows black hole mass as a function of jet power. Different
symbols correspond to blazars of different classes. Pearson
product-moment analysis (hereafter called Pearson analysis) is
applied to analyze the correlations between black hole mass and jet
power for all blazars. The results show that the correlations
between black hole mass and jet power for both Fermi blazars and
non-Fermi blazars are significant (Fermi blazars: number of points
$N=91$, significance level $P=3\times10^{-4}$, coefficient of
correlation $r=0.37$; non-Fermi blazars: $N=96$, $P=5\times10^{-3}$,
$r=0.28$).

\begin{figure}
\includegraphics[width=95mm, height=70mm]{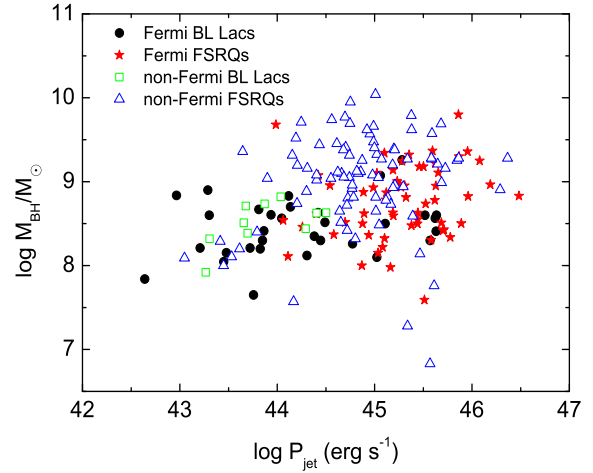}
\caption{Black hole mass as a function of jet power of various
classes. Fermi BL Lacs: black filled circles; Fermi FSRQs: red
filled stars; non-Fermi BL Lacs: green empty squares; non-Fermi
FSRQs: blue empty triangles. The uncertainties of jet kinetic power
and black hole mass are 0.7 dex and 1 dex respectively.}
\label{figure 12}
\end{figure}

\begin{figure}
\includegraphics[width=95mm, height=70mm]{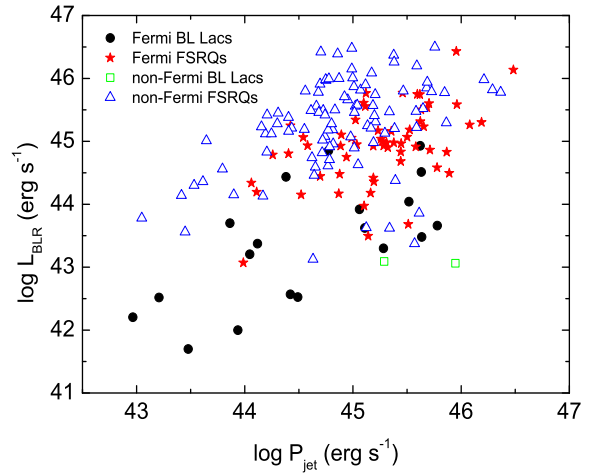}
\caption{Broad line luminosity as a function of jet power of various
classes. The meanings of different symbols are as same as Fig. 8.
The uncertainties of jet kinetic power and broad line luminosity are
0.7 dex and 0.5 dex respectively.} \label{figure 13}
\end{figure}

\begin{figure}
\includegraphics[width=95mm, height=70mm]{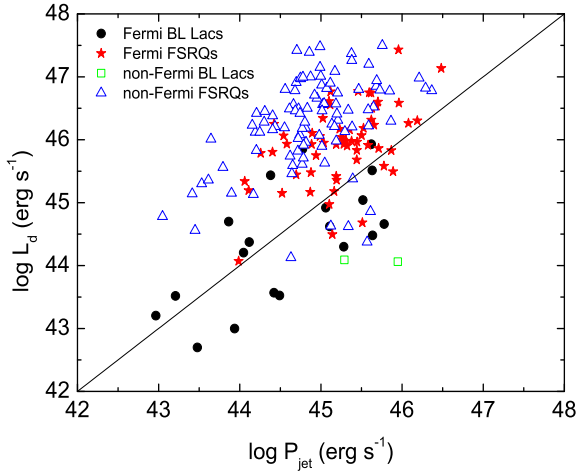}
\caption{Disk luminosity as a function of jet power of various
classes. The black line stands for $P_{\rm jet}=L_{\rm d}$. The
meanings of different symbols are as same as Fig. 8. The
uncertainties of jet kinetic power and disk luminosity are 0.7 dex
and 1 dex respectively.} \label{figure 14}
\end{figure}

\subsection{Broad line luminosity and disk luminosity vs jet power}

Figure 9 shows broad line luminosity as a function of jet power. The
results of Pearson analysis show that there are significant
correlations between broad line luminosity and jet power for Fermi
blazars and non-Fermi blazars ($N=78$, $P=2.6\times10^{-11}$,
$r=0.67$; $N=97$, $P=2\times10^{-3}$, $r=0.31$). Linear regression
is applied to analyze the correlation between broad line luminosity
and jet power. And we obtain $\log L_{\rm
BLR}\sim(0.94\pm0.12)logP_{jet}$ for Fermi blazars; $\log
L_{BLR}\sim(0.39\pm0.12)\log P_{\rm jet}$ for non-Fermi blazars
(95\% confidence level and $r=0.67, 0.31$). The Analysis of Variance
(ANOVA) is used to test the results of linear regression which shows
that it is valid for the results of linear regression (value
$F=60.89$, probability $P=2.6\times10^{-11}; F=10.13,
P=2\times10^{-3})$. Figure 10 shows disk luminosity as a function of
jet power. From Fig. 10, it is seen that compared with non-Fermi
blazars, Fermi blazars are closer to the $P_{\rm jet}=L_{\rm d}$;
the jet powers of some BL Lacs are larger than the disk luminosity
while some are opposite; the jet power is much smaller than the disk
luminosity for most of FSRQs.

We use multiple linear regression analysis to get the relationships
between the jet power and both the Eddington luminosity and the
broad line region luminosity for Fermi and non-Fermi blazars with
95\% confidence level and $r=0.7, 0.5$:
\begin{equation}
\log P_{\rm jet}=0.49(\pm0.06)\log L_{\rm BLR}-0.01(\pm0.16)\log
L_{\rm Edd}+23.6(\pm7),
\end{equation}

\begin{equation}
\log P_{\rm jet}=0.7(\pm0.14)\log L_{\rm BLR}-0.46(\pm0.17)\log
L_{\rm Edd}+34.71(\pm4.9).
\end{equation}
The ANOVA shows that it is valid for the results of multiple linear
regression ($F=33.5, P=5.13\times10^{-11}; F=13.8,
P=6.7\times10^{-6})$. From Equations (2) and (3), we see that for
Fermi blazars, the black hole mass does not have significant
influence on jet power while for non-Fermi blazars, both accretion
disk luminosity (referring to accretion rate; Sbarrato et al. 2014;
Ghisellini et al. 2014) and black hole mass have contributions to
the jet power.

\subsection{The blazar sequence}

The $\alpha_{\rm ro}$--$\alpha_{\rm ox}$ plot of our sample is shown
in Fig. 11. The $\alpha_{\rm ro}$--$\alpha_{\rm ox}$ plot of BZCAT
catalog is given in Fig. 27 of Abdo et al. (2010c). From Fig. 11 of
our sample, it is shown that FSRQs are exclusively located along the
top-left/bottom-right band; the $\alpha_{\rm ro}$ of most of BL Lacs
are located between 0 and 0.8, and 0.8--2 for $\alpha_{\rm ox}$. As
a comparison, it is found that the space of $0.5<\alpha_{\rm
ox}<0.8$ in Fig. 27 of Abdo et al. (2010c) has some BL Lacs while do
not in our Fig. 11; for both our Fig. 11 and Fig. 27 of Abdo et al.
(2010c), the top-right space of $\alpha_{\rm ro}$--$\alpha_{\rm ox}$
plot is empty; for the rest of space, our Fig. 11 is consistent with
Fig. 27 of Abdo et al. (2010c). Selection effects may cause the
deficiency of BL Lacs located in $0.5<\alpha_{\rm ox}<0.8$ in our
Fig. 11 because these BL Lacs are more likely to have nothing data
about redshift, black hole mass, jet power. Padovani \& Giommi
(1995) and Abdo et al. (2010c) presented that the $\alpha_{\rm ro}$
of HBL sources are located between 0.2 and 0.4, and 0.9--1.3 for
$\alpha_{\rm ox}$. In our sample, this region mainly includes
non-Fermi BL Lacs. Compared with non-Fermi FSRQs, Fermi FSRQs are
more located in top-left of Fig. 11.

Figure 12 shows synchrotron peak luminosity $L_{\rm peak}$ versus
synchrotron peak frequency $\nu_{\rm peak}$ for blazars and
Fermi-detected narrow-line Seyfert 1 galaxy. The synchrotron peak
luminosity and synchrotron peak frequency for four Fermi-detected
narrow-line Seyfert 1 galaxies are collected from Abdo et al.
(2009b). Making use of Pearson analysis, we find that there are
strong anti-correlations between $L_{\rm peak}$ and $\nu_{\rm peak}$
for all blazars, Fermi blazars and non-Fermi blazars respectively
($r=-0.637, P=4.9\times10^{-82}, N=711$; $r=-0.564,
P=1.1\times10^{-21}, N=242$; $r=-0.657, P=3.4\times10^{-59},
N=469$). In Fig. 12, we also find that the four Fermi-detected
narrow-line Seyfert 1 galaxies mix in the area of blazars populated,
and are in low $\nu_{\rm peak}$--low $L_{\rm peak}$ region, which
support that $\gamma$-loud narrow-line Seyfert 1 galaxies have
similar mechanisms with blazars, and that the low $\nu_{\rm
peak}$--low $L_{\rm peak}$ blazars are more likely to have low black
hole mass (see Chen \& Bai 2011).

Ghisellini \& Tavecchio (2008) extended the shape of SED from a
one-parameter (observed bolometric luminosity) to a two-parameter
sequence (black hole mass and accretion rate). So in Fig. 13, we
give synchrotron peak frequency ($\nu_{\rm peak}$) versus black hole
mass and Eddington ratio. From results of Pearson analysis, we get
that there is not significant correlation between $\nu_{\rm peak}$
and black hole mass for all blazars ($r=0.03, P=0.38$) while there
is significant anti-correlation between $\nu_{\rm peak}$ and
Eddington ratio for all blazars ($r=-0.22, P=3.8\times10^{-7}$).

\begin{figure}
\includegraphics[width=95mm, height=70mm]{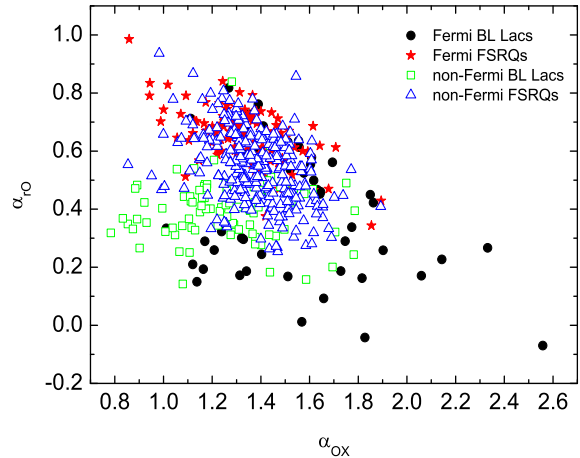}
\caption{The $\alpha_{\rm ro}$--$\alpha_{\rm ox}$ plot. The meanings
of different symbols are as same as Fig. 8.} \label{figure 14}
\end{figure}

\begin{figure}
\includegraphics[width=95mm, height=70mm]{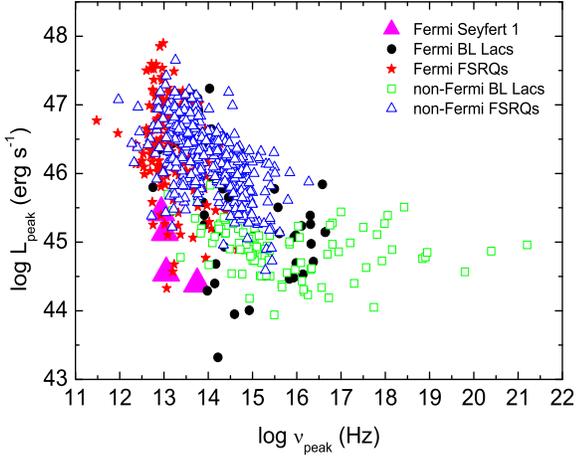}
\caption{Synchrotron peak luminosity $L_{\rm peak}$ versus
synchrotron peak frequency $\nu_{\rm peak}$ for blazars, and
Fermi-detected narrow-line Seyfert 1 galaxies. The meanings of
different symbols of blazars are as same as Fig. 8. Fermi-detected
narrow-line Seyfert 1 galaxies: magenta filled triangles.}
\label{figure 16}
\end{figure}

\begin{figure}
\includegraphics[width=160mm, height=120mm]{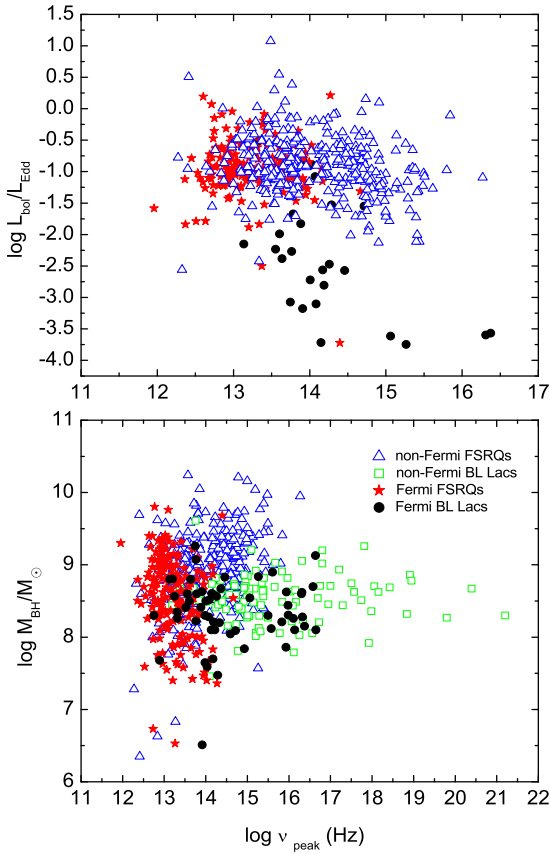}
\caption{Black hole mass (bottom panel) and Eddington ratio (top
panel) versus synchrotron peak frequency ($\nu_{\rm peak}$). The
meanings of different symbols of blazars are as same as Fig. 8.}
\label{figure 17}
\end{figure}

\section{Discussions and conclusions}

\subsection{Possible biases in the evaluations of sample and parameters}

For our sample, the selection criteria are: (i) we tried to select
the largest group of blazars included in BZCAT with reliable broad
line luminosity, redshift, black hole mass and jet kinetic power;
(ii) in order to have reliable sample, we did not consider the
candidate blazars of unknown type; (iii) the non-Fermi blazars,
which were detected by EGRET or recorded in 1LAC but missed in 2LAC,
were not included in our sample; (iv) the Fermi blazars only focused
on clean 2LAC sample. Based on the above criteria, the following
objects can be missed: blazars without measured redshift (mainly
including BL Lacs), blazars without measured black hole mass,
extended 1.4 GHz radio data and broad line data, blazars of
uncertain type (BZU), blazars which were detected by EGRET or
recorded in 1LAC but missed in 2LAC and blazars classed into
non-clean 2LAC. From our selection criteria of sample, it is shown
that the number of BL Lacs is missed more in our sample than that of
FSRQs because compared with FSRQs, BL Lacs have much less
information about redshift, black hole mass and broad line data. Via
comparing our redshift distribution to redshift distribution of
complete BZCAT sample, we find that the whole distribution and mean
value of redshift from our sample agree with the results of complete
BZCAT sample. From complete 2LAC sample (Fig. 12 of Ackermann et al.
(2011)), it is shown that the redshift distribution peaks around
$z=1$ for FSRQs and extends to $z=3.1$; the redshift distribution
for BL Lacs extends to $z=1.5$ and shows a broad maximum between
$0.1$ and $0.2$. So through comparing redshift distributions between
our Fermi sample (Fig. 1) and complete 2LAC sample, we find that our
redshift distribution of Fermi blazars is similar to redshift
distribution of complete 2LAC sample. However, this does not
necessarily indicate that the distributions of all other quantities
are also similar.

In order to reduce the uncertainty, we tried to select the data from
a same paper and/or a uniform method. Firstly, for most of FSRQs
from our sample (529 FSRQs), black hole masses were estimated by
traditional virial method; for most of BL Lacs (150 BL Lacs), the
black hole masses can be estimated from the properties of their host
galaxies with either $M_{\rm BH}-\sigma$ or $M_{\rm BH}-L$
relations, where $\sigma$ and $L$ are the stellar velocity
dispersion and the bulge luminosity of the host galaxies. For a few
blazars (29 blazars), the BH masses were estimated from minimum
timescale for flux variations ($M_{\rm BH}-\triangle t_{\rm min}$
from Xie et al. (2004)). In our sample, we found that for same
blazars, black hole masses estimated from different emission line or
different authors had a little difference. For the blazars, when
more than one black hole masses were got, we used average BH mass
instead. Moreover, we should be caution in the black hole masses
estimated from virial method or relations of host galaxies because
of non-thermal dominance for blazars and contamination of host
galaxy light. Secondly, for calculating the total luminosity of the
broad lines, we and the other authors adopted method of Celotti,
Padovani \& Ghisellini (1997) which scaled several strong emission
lines to the quasar template spectrum of Francis et al. (1991). Via
comparing broad-line luminosities estimated by us and the other
authors, we found that for some blazars, the broad-line luminosities
between us and the other authors were slightly different. The
possible reasons were that we and the other authors used different
lines to calculate broad-line luminosity; variability also can cause
the difference of broad-line luminosity. In these blazars, we used
average broad-line luminosity instead. It also is worth noting that
BLR luminosity is not a direct measure of the disc luminosity and
furthermore that it is possible that BL Lacs may have a less
luminous accretion disc (radiatively inefficient accretion flow)
than FSRQs which would make it difficult to estimate the accretion
rate from the BLR or disc luminosity. Thirdly, the jet power of our
sample was estimated from extended radio luminosity at 300 MHz.
Following Meyer et al. (2011), we extrapolated 1.4 GHz extended
radio emission to 300 MHz luminosity, using a low frequency index of
$\alpha=1.2$ ($F_{\rm \nu}\propto\nu^{-\alpha}$). It was possible
that the extrapolation can bring in uncertainty for estimating jet
power. The uncertainty in jet power was dominated by the scatter in
the correlation of Cavagnolo et al. (2010). In XZ14, we have
compared the jet cavity power to jet power from modeling the SED of
Ghisellini et al. (2010), and found that on average, the jet power
from Ghisellini et al. (2010) was slightly larger than the jet
cavity power. The possible reasons are as follows (Kang et al. 2014;
Kharb et al. 2010): (i) when Ghisellini et al. (2010) fitted the
SED, they assumed one proton per emitting electron in the jet,
whereas the jet power from modeling the SED would be reduced if the
jet also included a fraction of $e^\pm$ pairs; (ii) the extended
radio luminosity is affected by interacting with the environment on
kiloparsec-scales. Finally, we discussed $L_{\rm peak}$ and
$\nu_{\rm peak}$. At present, fitting simultaneous SED data from
long time is an ideal method to determine $L_{\rm peak}$ and
$\nu_{\rm peak}$. However, it is hard to obtain simultaneous
multi-wave band data for a large sample. For most of our sample, we
and Finke (2011) used the empirical relationship of Abdo et al.
(2010c) to estimate $\nu_{\rm peak}$. Abdo et al. (2010c) have
pointed out that their method assumed that the optical and X-ray
fluxes are not contaminated by thermal emission from the disk or
accretion. In blazars where thermal flux components are not
negligible (this should probably occur more frequently in low radio
luminosity sources) the method may lead to a significant
overestimation of the position of $\nu_{\rm peak}$. For blazars
without X-ray flux, we adopted $\alpha_{\rm ro}$ to estimate
$\nu_{\rm peak}$. The $\nu_{\rm peak}$ and $L_{\rm peak}$ of
Nieppola et al. (2006) were estimated by fitting non-simultaneous
SED data. The $\nu_{\rm peak}$ and $L_{\rm peak}$ of Meyer et al.
(2011) were estimated by fitting non-simultaneous average SED. For
many blazars, we also used radio luminosity at 5 GHz to estimate the
synchrotron peak luminosity. As we know, blazars are high
variability. Therefore, the above factors or processes can bring in
uncertainties for estimating $\nu_{\rm peak}$ and $L_{\rm peak}$.
For $R_{\rm c}$ and $R_{\rm v}$, variability can lead to
uncertainties.

\subsection{The basic properties of Fermi blazars}

By comparing the main parameters between Fermi blazars and non-Fermi
blazars, we obtain the below results. (i) The redshift distributions
between Fermi blazars and non-Fermi blazars have significant
difference. For all blazars and FSRQs, Fermi sources have smaller
mean redshift than non-Fermi sources while for only BL Lacs, the
result is the opposite. So compared with non-Fermi FSRQs, Fermi
FSRQs are relatively nearby objects because the $\gamma$-ray of
greater distance object is likely to be absorbed (Piner et al.
2008). But for BL Lacs, the opposite result can be explained as
follows. The result of comparison may be a selection effect because
based on our sample selection criteria, many BL Lacs are missed. In
addition, another possible explanation is that redshift may not be
as a main factor for difference between Fermi and non-Fermi BL Lacs,
e.g. if the beaming effect is a main reason for difference between
Fermi and non-Fermi BL Lacs, a BL Lac with small redshift still can
be as a Fermi source as long as it has enough large beaming factor.
(ii) The black hole mass distributions between Fermi blazars and
non-Fermi blazars have significant difference. Fermi blazars have
smaller mean black hole mass than non-Fermi blazars. Generally, one
may consider Fermi blazars with a larger black hole mass because a
blazar with larger luminosity may have a larger black hole mass
(Ghisellini et al. 2010). Our results seem to contradict with the
idea. Meier (1999) has demonstrated explicitly that it is not
necessary to have a relatively massive black hole mass to produce
powerful jet. Based on current accretion and jet production theory
(e.g. Blandford \& Znajek 1977; Meier 1999), jet power is tied to
the spinning of black hole. Then, it is quite possible to have
highly powered jet with small black hole mass if the black hole has
a high spin (the maximum jet power is close to the Eddington
luminosity or an efficiency of $\sim$140\% of the accretion power
from Tchekhovskoy et al. (2011)). The other possibility is that
black hole mass is not main factor for difference between Fermi
blazars and non-Fermi blazars. (iii) The jet power distributions
between Fermi blazars and non-Fermi blazars have significant
difference. Fermi blazars have larger mean jet power than non-Fermi
blazars. In our sample, the jet power estimated requires assumption
about the energy required to inflate the lobe, and is free of
beaming effect because our jet power is estimated from extended
radio emission. So the result supports that Fermi blazars are likely
to have a more powerful jet. (iv) The Eddington ratio distributions
between Fermi FSRQs and non-Fermi FSRQs do not have significant
difference, which does not mean that the accretion rate
distributions between them also are similar. The fact that similar
Eddington ratio distributions between them means similar accretion
rate distributions is only implied if the black hole masses are also
very similar. The broad-line luminosity distributions between Fermi
FSRQs and non-Fermi FSRQs have significant difference. Compared with
non-Fermi FSRQs, Fermi FSRQs have smaller mean broad-line
luminosity. Theoretically, the continuum flux is believed to be
responsible for ionizing the cloud material in the broad line region
(e.g. Arshakian et al. 2010). So compared with non-Fermi blazars,
Fermi blazars may have larger broad line luminosity because Fermi
blazars have larger jet power than non-Fermi blazars. But this is
opposite with our results. A probable explanation is that broad-line
luminosity is not main factor for difference between Fermi blazars
and non-Fermi blazars, e.g. a blazar with low broad line luminosity
still can be as Fermi source as long as it has enough large beaming
factor. Moreover, it is possible due to redshift distribution
effect: Fermi FSRQs have smaller mean redshift than non-Fermi FSRQs,
and increasing average redshift may cause increasing line luminosity
(see Ghisellini \& Tavecchio 2015). (v) The $R_{\rm c}$ can be as
indicators of Doppler beaming and orientation. From our sample, we
find that the mean difference between $R_{\rm c}$ for Fermi and
non-Fermi blazars is small and not significant. Kharb et al. (2010)
found that the $R_{\rm v}$ appears to be a better indicator of
orientation than the traditionally used $R_{\rm c}$, since the
optical luminosity is likely to be a better measure of intrinsic jet
power (e.g. Maraschi et al. 2008; Ghisellini et al. 2009c) than
extended radio luminosity. This is due to the fact that the optical
continuum luminosity is correlated with the emission-line luminosity
over four orders of magnitude (Yee \& Oke 1978), and the
emission-line luminosity is tightly correlated with the total jet
kinetic power (Rawlings \& Saunders 1991). The extended radio
luminosity, on the other hand, is suggested to be affected by
interaction with the environment on kiloparsec-scales. Our comparing
results show that the $R_{\rm v}$ distributions between Fermi
blazars and non-Fermi blazars are significant difference and that
the mean value of $R_{\rm v}$ for Fermi blazars is larger than that
for non-Fermi blazars. Therefore, our results provide new supports
for that compared with non-Fermi blazars, Fermi blazars have a
stronger beaming effect.

The jet formation remains one of the unsolved fundamental problems
in astrophysics (e.g. Meier, Koide \& Uchida 2001). Many models have
been proposed to explain the origin of jets. In current theoretical
models of the formation of jet, power is generated via accreting
material and extraction of rotational energy of disc/black hole
(Blandford \& Znajek 1977; Blandford \& Payne 1982), and then
converted into the kinetic power of the jet. Our results show that
the correlations between black hole mass and jet power for both
Fermi and non-Fermi blazars are significant. The results of Pearson
analysis show that there are significant correlations between broad
line luminosity and jet power for both Fermi blazars and non-Fermi
blazars, which supports that the jet power has a close link with
accretion rate (Sbarrato et al. 2014; Ghisellini et al. 2014). The
result is consistent with other authors (e.g. Rawlings \& Saunders
1991; Falcke \& Biermann 1995; Serjeant, Rawlings \& Maddox 1998;
Cao \& Jiang 1999; Wang, Luo \& Ho 2004; Liu, Jiang \& Gu 2006; Xie
et al. 2007; Ghisellini, Tavecchio \& Ghirlanda 2009a; Ghisellini et
al. 2009b, 2010, 2011; Gu, Cao \& Jiang 2009; Sbarrato et al. 2012).
Linear regression is applied to analyze the correlation between
broad line luminosity and jet power, and we obtain $\log L_{\rm
BLR}\sim(0.94\pm0.12)\log P_{\rm jet}$ for Fermi blazars and $\log
L_{\rm BLR}\sim(0.39\pm0.12)\log P_{\rm jet}$ for non-Fermi blazars.
In addition, from Equations (2) and (3), we find that for Fermi
blazars, the black hole mass does not have significant influence on
jet power while for non-Fermi blazars, both accretion disk
luminosity (or accretion rate) and black hole mass have
contributions to the jet power. These also can explain our result
that the distributions of jet power between Fermi and non-Fermi
blazars are significant difference.

\subsection{The blazar sequence}

From Fig. 12 and making use of Pearson analysis, we find that there
are strong anti-correlations between $L_{\rm peak}$ and $\nu_{\rm
peak}$ for all blazars, Fermi blazars and non-Fermi blazars
respectively. The results support the ``blazar sequence'' and are
not same with results of Meyer et al. (2011). In Fig. 4 of Meyer et
al. (2011), an ``L''-shape in the $L_{\rm peak}$--$\nu_{\rm peak}$
plot seems to have emerged which destroys the ``blazar sequence''.
However, the ``L''-shape in Meyer et al. (2011) may be a selection
effect of sample. A very possible reason is that in $10^{14} {\rm
Hz}<\nu_{\rm peak}<10^{15} {\rm Hz}$ interval of $L_{\rm
peak}$--$\nu_{\rm peak}$ plot of Meyer et al. (2011), there are few
blazars. If the interval is full of some blazars, then the
``L''-shape will be weaken or disappeared. In our Fig. 12, the
interval has many blazars which fill in the interval of Fig. 4 of
Meyer et al. (2011). And the result is consistent with others. From
Fig. 2 of Nieppola et al. (2008) and Fig. 2, 5 of Finke (2013), it
was seen that their the interval still has some blazars. Giommi et
al. (2012) used extensive Monte Carlo simulations to get the
relation plot of $\nu_{\rm peak}-L_{\rm 5GHz}$ which shows that the
interval still has many blazars. The Compton dominance, the ratio of
the peak of the Compton to the synchrotron peak luminosities, is
essentially a redshift-independent quantity. Finke (2013) used
Compton dominance to study the ``blazar sequence''. The results of
Finke (2013) showed that a correlation exists between Compton
dominance and the peak frequency of the synchrotron component for
all blazars, including ones with unknown redshift. The sample of
Finke (2013) is the 2LAC clean sample. The $\gamma$-ray flares in
blazars are likely to occur when the sources are in the high state
(Fan et al. 1998). If this is true, the blazars from Finke (2013)
are likely to stand for active states of the sources, not their
necessarily averaged status. Finke (2013) also presented that SSC
emission has the same beaming pattern as synchrotron, and thus
Compton dominance does not depend on the viewing angle; however, EC
emission does not have the same beaming pattern as synchrotron and
SSC, and so Compton dominance is dependent on the viewing angle.
Piner \& Edwards (2013) have found that beaming factors in the
$\gamma$-ray and radio bands are different yet correlated for TeV
blazars. An obvious explanation for the `bulk Lorentz factor crisis'
is that the radio and $\gamma$-ray emissions are produced in
different parts of the jet with different bulk Lorentz factors
(Henri \& Sauge 2006; Piner, Pant \& Edwards 2008, 2010; Piner \&
Edwards 2013). Therefore, it is possible that the correlation
between Compton dominance and peak frequency of the synchrotron
component can be resulted from beaming effect. For our Fig. 12, if
high $\nu_{\rm peak}$ high $L_{\rm peak}$ blazars are included, the
correlation of $\nu_{\rm peak}-L_{\rm peak}$ will be destroyed. The
results of Giommi et al. (2012) also showed that the
phenomenological ``blazar sequence'' is a selection effect. Nieppola
et al. (2008) proposed that after being Doppler-corrected, the
anti-correlation between $\nu_{\rm peak}$ and $L_{\rm peak}$ become
positive correlation. However, due to limit of Doppler factor, we
can not determine the effects of beaming on the blazar sequence in
this paper. Ghisellini \& Tavecchio (2008) revisited the blazar
sequence and proposed that the power of the jet and SED of its
emission are linked to the mass of black hole and the accretion
rate. From results of Pearson analysis, we get that there is not
significant correlation between synchrotron peak frequency and black
hole mass for all blazars while significant anti-correlation between
synchrotron peak frequency and Eddington ratio for all blazars. The
scatter in $\nu_{\rm peak}-L_{\rm bol}/L_{\rm Edd}$ plot can be
resulted from uncertainties of $\nu_{\rm peak}$ and black hole mass.

\section*{Acknowledgments}

We sincerely thank anonymous referee for valuable comments and
suggestions. DRX also thanks Minfeng Gu for helpful suggestions.
This work is financially supported by the National Nature Science
Foundation of China (11163007, U1231203, 11063004, 11133006 and
11361140347) and the Strategic Priority Research Program ``The
emergence of Cosmological Structures'' of the Chinese Academy of
Sciences (grant No. XDB09000000). This research has made use of the
NASA/IPAC Extragalactic Database (NED), that is operated by Jet
Propulsion Laboratory, California Institute of Technology, under
contract with the National Aeronautics and Space Administration.

\begin{table*}
 \centering
  \caption{The non-Fermi sample.}
  \begin{tabular}{@{}llrrrlrlr@{}}
  \hline\hline
   BZCAT name  &   Other name  &   RA  &   Redshift    &   $\log\nu_{\rm peak}$ &   $\log P_{\rm jet}$    &   $\log M_{\rm BH}$ &   $\log L_{\rm BLR}$    &   $\log R_{\rm c}$   \\
    &       &   DEC &       &   $\log L_{\rm peak}$   &       &   ref &   ref &   ref \\
(1) &   (2) &   (3) &   (4) &   (5) &   (6) &   (7) &   (8) &   (9) \\
 \hline
BZQ J0006-0623  &   0003-066    &   00 06 13.9  &   0.347   &   13.16$^\dagger$ &   44.63   &       &   43.13   &   1.48    \\
    &       &   -06 23 35   &       &   46.22$^\dagger$ &    &       &   C99 &   C07 \\
BZQ J0010+1058  &   0007+106    &   00 10 31.0  &   0.089   &   14.88   &   43.41    &   8.29    &   44.14   &   0.62    \\
    &       &   +10 58 30   &       &   45.05   &    &   C12 &   C12 &   K10 \\
BZQ J0017+8135  &   0014+813    &   00 17 08.5  &   3.366   &   14.55$^\dagger$ &      &       &   46.62   &       \\
    &       &   +81 35 08   &       &   46.99   &      &       &   C99 &       \\
BZQ J0019+7327  &   0016+731    &   00 19 45.8  &   1.781   &   13.26$^\dagger$ &   45.19   &   8.93    &   44.98   &   1.27    \\
    &       &   +73 27 30   &       &   47.65$^\dagger$ &    &   C12 &   C12 &   K10 \\
BZB J0022+0006  &   SDSS J002200.95+000657.9    &   00 22 00.9  &   0.306   &   16.17   &       &   8.49    &       &       \\
    &       &   +00 06 58   &       &   44.28   &      &   P11 &       &       \\
BZQ J0038+4137  &   0035+413    &   00 38 24.8  &   1.353   &   13.20   &      &   8.53    &   44.64   &       \\
    &       &   +41 37 06   &       &   46.75   &      &   C12 &   C12 &       \\
BZB J0056-0936  &   SDSS J005620.07-093629.7    &   00 56 20.1  &   0.103   &   15.01   &       &   8.89,8.39   &       &       \\
    &       &   -09 36 30   &       &   44.82   &      &   L11 &       &       \\
BZQ J0057-0024  &   SDSS J005716.99-002433.2    &   00 57 17.0  &   2.752   &   13.99   &    &   9.73    &   45.68   &       \\
    &       &   -00 24 33   &       &   46.62   &   &   S11 &   S11 &       \\
BZQ J0059+0006  &   0056-001    &   00 59 05.5  &   0.719   &   13.58   &    &   8.71,8.37,9.03,8.86 &   44.91   &       \\
    &       &   +00 06 52   &       &   46.49   &    &   W02,L06,S11 &   L06 &       \\
BZB J0110+4149  &   NPM1G +41.0022  &   01 10 04.8  &   0.096   &   17.74$^\dagger$ &       &   8.51    &       &       \\
    &       &   +41 49 51   &       &   44.05$^\dagger$ &    &   W09 &       &       \\
BZQ J0115-0127  &   0112-017    &   01 15 17.1  &   1.365   &   13.54   &   &   7.85    &   45.26   &       \\
    &       &   -01 27 05   &       &   46.85   &     &   C12 &   C12 &       \\
BZQ J0121+1149  &   0119+115    &   01 21 41.6  &   0.57    &   12.81   &   45.12   &       &   43.63   &   0.84    \\
    &       &   +11 49 50   &       &   46.29   &    &       &   C99 &   K10 \\
\hline
\end{tabular}
\begin{quote}
(a) References. A85: Antonucci \& Ulvestad (1985); Ca99: Cassaro et
al. (1999); C12: Chai et al. (2012); C07: Cooper et al. (2007); C04:
Caccianiga \& Marcha (2004); C99: Cao \& Jiang (1999); F03: Falomo
et al. (2003); K10: Kharb et al. (2010); L08: Landt \& Bignall
(2008); L06: Liu et al. (2006); L11: Leon-Tavares et al. (2011);
M93: Murphy et al. (1993); P11: Plotkin et al. (2011); S11: Shen et
al. (2011); W09: Wu et al. (2009); W05: Woo et al. (2005); W04: Wang
et al. (2004); W02: Woo \& Urry (2002); W00: White et al. (2000).

(b) For $\log\nu_{\rm peak}$ and $\log L_{\rm peak}$, values with a
`` $\dagger$ '' represent that they are directly from references.

(c) This table is published in its entirety in the electronic
edition. A portion is shown here for guidance.
\end{quote}
\end{table*}


\begin{thebibliography}{99}
\bibitem[\protect\citeauthoryear{Aatrokoski et al.}{2011}]{b1} Aatrokoski J. et al., 2011, A\&A, 536, A15
\bibitem[\protect\citeauthoryear{Abdo et al.}{2009}]{b1} Abdo A.A., Ackermann M., Ajello M. et al., 2009a, ApJ, 700, 597
\bibitem[\protect\citeauthoryear{Abdo et al.}{2009}]{b1} Abdo A.A. et al., 2009b, 707, L142
\bibitem[\protect\citeauthoryear{Abdo et al.}{2010a}]{b2} Abdo A.A., Ackermann M., Ajello M. et al., 2010a, ApJS, 188, 405
\bibitem[\protect\citeauthoryear{Abdo et al.}{2010b}]{b3} Abdo A.A., Ackermann M., Ajello M. et al., 2010b, ApJ, 715, 429
\bibitem[\protect\citeauthoryear{Abdo et al.}{2010c}]{b4} Abdo A.A., Ackermann M., Agudo I. et al., 2010c, ApJ, 716, 30
\bibitem[\protect\citeauthoryear{Abdo et al.}{2012}]{b6} Abdo A.A., Ackermann M., Ajello M. et al., 2012, ApJS, 199, 31
\bibitem[\protect\citeauthoryear{Ackermann et al.}{2011a}]{b7} Ackermann M., Ajello M., Allafort A. et al., 2011, ApJ, 743, 171
\bibitem[\protect\citeauthoryear{Angel}{1980}]{b9} Angel J.R.P. \& Stockman H.S., 1980, ARA\&A, 18, 321
\bibitem[\protect\citeauthoryear{Antonucci}{1985}]{b9} Antonucci R.R.J. \& Ulvestad J.S., 1985, ApJ, 294, 158
\bibitem[\protect\citeauthoryear{Arshakian}{2010}]{b9} Arshakian T.G., Leon-Tavares J., Lobanov A.P., Chavushyan V.H., Shapovalova A.I., Burenkov A.N. \& Zensus J.A., 2010, MNRAS, 401, 1231
\bibitem[\protect\citeauthoryear{Balmaverde et al.}{2008}]{b10} Blandford R.D. \& Znajek R.L., 1977, MNRAS, 179, 433
\bibitem[\protect\citeauthoryear{Balmaverde et al.}{2008}]{b10} Blandford R.D. \& Payne D.G., 1982, MNRAS, 199, 883
\bibitem[\protect\citeauthoryear{Caccianiga}{2004}]{b10} Caccianiga A. \& Marcha M.J.M., 2004, MNRAS, 348, 937
\bibitem[\protect\citeauthoryear{Cassaro}{1999}]{b10} Cassaro P. et al., 1999, A\&AS, 139, 601
\bibitem[\protect\citeauthoryear{Cavagnolo}{2010}]{b13} Cavagnolo K.W. et al., 2010, ApJ, 720, 1066
\bibitem[\protect\citeauthoryear{Cavaliere \& D'Elia}{2002}]{b13} Cavaliere A. \& D'Elia V., 2002, ApJ, 571, 226
\bibitem[\protect\citeauthoryear{Cao \& Jiang}{1999}]{b14} Cao X. \& Jiang D.R., 1999, MNRAS, 307, 802
\bibitem[\protect\citeauthoryear{Celotti et al.}{1997}]{b16} Celotti A., Padovani P. \& Ghisellini G., 1997, MNRAS, 286, 415
\bibitem[\protect\citeauthoryear{Chai et al.}{2012}]{b19} Chai B., Cao X. \& Gu M., 2012, ApJ, 759, 114
\bibitem[\protect\citeauthoryear{Chen  et al.}{2011}]{b19} Chen L. \& Bai J.M., 2011, ApJ, 735, 108
\bibitem[\protect\citeauthoryear{Cheng \& Zhang}{2000}]{b21} Cheng K.S., Zhang X. \& Zhang L., 2000, ApJ, 537, 80
\bibitem[\protect\citeauthoryear{Cooper}{2007}]{b21} Cooper N.J., Lister M.L. \& Kochanczyk M.D., 2007, ApJS, 171, 376
\bibitem[\protect\citeauthoryear{Falcke \& Biermann}{1995}]{b23} Falcke H. \& Biermann P.L., 1995, A\&A, 293, 665
\bibitem[\protect\citeauthoryear{Falomo}{2003}]{b23} Falomo R., Kotilainen J.K., Carangelo N. \& Treves A., 2003, ApJ, 595, 624
\bibitem[\protect\citeauthoryear{Fan}{1998}]{b23} Fan J.H. et al., 1998, A\&A, 338, 27
\bibitem[\protect\citeauthoryear{Finke}{1994}]{b24} Finke J.D., 2013, ApJ, 763, 134
\bibitem[\protect\citeauthoryear{Fossati et al.}{1998}]{b24} Fossati G., Maraschi L., Celotti A. et al., 1998, MNRAS, 299, 433
\bibitem[\protect\citeauthoryear{Francis et al.}{1991}]{b26} Francis P.J., Hewett P.C., Foltz C.B., Chaffee F.H., Weymann R.J. \& Morris S.L., 1991, ApJ, 373, 465
\bibitem[\protect\citeauthoryear{Francis et al.}{1991}]{b26} Georganopoulos M., Kirk J.G. \& Mastichiadis A., 2001, in ASP Conf. Ser. 227, Blazar Demographics and Physics, ed. P. Padovani \& C. Megan Urry (San Francisco, CA: ASP), 116
\bibitem[\protect\citeauthoryear{Ghisellini}{1998}]{b28} Ghisellini G., Celotti A., Fossati G. et al., 1998, MNRAS, 301, 451
\bibitem[\protect\citeauthoryear{Ghisellini \& Tavecchio}{2008}]{b32} Ghisellini G. \& Tavecchio F., 2008, MNRAS, 387, 1669
\bibitem[\protect\citeauthoryear{Ghisellini et al.}{2009a}]{b33} Ghisellini G., Maraschi L. \& Tavecchio F., 2009a, MNRAS, 396, 105
\bibitem[\protect\citeauthoryear{Ghisellini et al.}{2009b}]{b34} Ghisellini G., Tavecchio F., Foschini L., Ghirlanda G., Maraschi L. \& Celotti A., 2009b, MNRAS, 402, 497
\bibitem[\protect\citeauthoryear{Ghisellini et al.}{2009c}]{b34} Ghisellini G., Tavecchio F. \& Ghirlanda G., 2009c, MNRAS, 399, 2041
\bibitem[\protect\citeauthoryear{Ghisellini et al.}{2010}]{b37} Ghisellini G., Tavecchio F., Foschini L., Ghirlanda G., Maraschi L. \& Celotti A., 2010, MNRAS, 402, 497
\bibitem[\protect\citeauthoryear{Ghisellini et al.}{2011}]{b38} Ghisellini G., Tavecchio F., Foschini L. \& Ghirlanda G., 2011, MNRAS, 414, 2674
\bibitem[\protect\citeauthoryear{Ghisellini et al.}{2014}]{b38} Ghisellini G., Tavecchio F., Maraschi L., Celotti A. \& Sbarrato T., 2014, Nature, 515, 376
\bibitem[\protect\citeauthoryear{Ghisellini et al.}{2015}]{b38} Ghisellini G. \& Tavecchio F., 2015, MNRAS, 448, 1060
\bibitem[\protect\citeauthoryear{Giommi et al.}{1999}]{b38} Giommi P., Menna M.T. \& Padovani P., 1999, MNRAS, 310, 465
\bibitem[\protect\citeauthoryear{Giommi et al.}{2012}]{b38} Giommi P., Padovani P., Polenta G., Turriziani S., D'Elia V. \& Piranomonte S., 2012, MNRAS, 420, 2899
\bibitem[\protect\citeauthoryear{Giommi et al.}{2013}]{b38} Giommi P., Padovani P. \& Polenta G., 2013, MNRAS, 431, 1914
\bibitem[\protect\citeauthoryear{Gu et al.}{2009}]{b39} Gu M., Cao X. \& Jiang D.R., 2009, MNRAS, 396, 984
\bibitem[\protect\citeauthoryear{Henri}{2006}]{b39} Henri G. \& Sauge L., 2006, ApJ, 640, 185
\bibitem[\protect\citeauthoryear{Jorstad et al.}{2001}]{b39} Jorstad S.G., Marscher A.P., Mattox J.R., Wehrle A.E., Bloom S.D. \& Yurchenko A.V., 2001, ApJS, 134, 181
\bibitem[\protect\citeauthoryear{Kang et al.}{2014}]{b39} Kang S.J., Chen L. \& Wu Q.W., 2014, ApJS, 215, 5
\bibitem[\protect\citeauthoryear{Kharb et al.}{2010}]{b39} Kharb P., Lister M.L. \& Cooper N.J., 2010, ApJ, 710, 764
\bibitem[\protect\citeauthoryear{Kovalev}{2009}]{b39} Kovalev Y.Y. et al., 2009, ApJ, 696, L17
\bibitem[\protect\citeauthoryear{Landt}{2008}]{b39} Landt H. \& Bignall H.E., 2008, MNRAS, 391, 967
\bibitem[\protect\citeauthoryear{Leon¨CTavares et al.}{2011a}]{b42} Leon-Tavares J., Valtaoja E., Chavushyan V.H. et al., 2011, MNRAS, 411, 1127
\bibitem[\protect\citeauthoryear{Linford et al.}{2011}]{b42} Linford J.D., Taylor G.B., Romani R. et al., 2011, ApJ, 726, 16
\bibitem[\protect\citeauthoryear{Linford et al.}{2012}]{b42} Linford J.D., Taylor G.B., Romani R. et al., 2012, ApJ, 744, 177
\bibitem[\protect\citeauthoryear{Lister et al.}{2009a}]{b42} Lister M.L. et al., 2009a, AJ, 138, 1874
\bibitem[\protect\citeauthoryear{Lister et al.}{2009b}]{b42} Lister M.L. et al., 2009b, ApJ, 696, 22
\bibitem[\protect\citeauthoryear{Liu et al.}{2006}]{b42} Liu Y., Jiang D.R. \& Gu M.F., 2006, ApJ, 637, 669
\bibitem[\protect\citeauthoryear{Madau}{1987}]{b43} Madau P., Ghisellini G. \& Persic M., 1987, MNRAS, 224, 257
\bibitem[\protect\citeauthoryear{Maraschi et al.}{2008}]{b46} Maraschi L., Foschini L., Ghisellini G., Tavecchio F. \& Sambruna R.M., 2008, MNRAS, 391, 1981
\bibitem[\protect\citeauthoryear{Maraschi et al.}{1992}]{b46} Maraschi L. \& Tavecchio F., 2003, ApJ, 593, 667
\bibitem[\protect\citeauthoryear{Massaro et al.}{2009}]{b46} Massaro E. et al., 2009, A\&A, 495, 691
\bibitem[\protect\citeauthoryear{Massaro et al.}{2012}]{b46} Massaro F., Paggi A., D'Abrusco R. \& Tosti G., 2012, ApJ, 750, L35
\bibitem[\protect\citeauthoryear{Mead}{1990}]{b47} Mead A.R.G., Ballard K.R., Brand P.W.J.L. et al., 1990, A\&AS, 83, 183
\bibitem[\protect\citeauthoryear{Meier et al.}{2001}]{b48} Meier D.L., Koide S. \& Uchida Y., 2001, Science, 291, 84
\bibitem[\protect\citeauthoryear{Meier et al.}{1999}]{b48} Meier D.L., 1999, ApJ, 522, 753
\bibitem[\protect\citeauthoryear{Meyer et al.}{2011}]{b48} Meyer E.T., Fossati G., Georganopoulos M. \& Lister M.L., 2011, ApJ, 740, 98
\bibitem[\protect\citeauthoryear{Murphy}{1993}]{b48} Murphy D.W., Browne I.W.A. \& Perley R.A., 1993, MNRAS, 264, 298
\bibitem[\protect\citeauthoryear{Nemmen et al.}{2001}]{b48} Nemmen R.S., Georganopoulos M., Guiriec S., Meyer E.T., Gehrels N., \& Sambruna R.M., 2012, Science, 338, 1445
\bibitem[\protect\citeauthoryear{Netzer et al.}{1990}]{b48} Netzer H., 1990, Active Galactic Nuclei, 57
\bibitem[\protect\citeauthoryear{Nieppola et al.}{2006}]{b48} Nieppola E., Tornikoski M. \& Valtaoja E., 2006, A\&A, 445, 441
\bibitem[\protect\citeauthoryear{Nieppola et al.}{2008}]{b48} Nieppola E., Valtaoja E., Tornikoski M., Hovatta T. \& Kotiranta M., 2008, A\&A, 488, 867
\bibitem[\protect\citeauthoryear{Orr}{1982}]{b48} Orr M.J.L. \& Browne I.W.A., 1982, MNRAS, 200, 1067
\bibitem[\protect\citeauthoryear{Padovani}{1992}]{b49} Padovani P. \& Giommi P., 1995, ApJ, 444, 567
\bibitem[\protect\citeauthoryear{Padovani}{2003}]{b49} Padovani P., Perlman E.S., Landt H., Giommi P. \& Perri M., 2003, ApJ, 588, 128
\bibitem[\protect\citeauthoryear{Padovani}{2007}]{b49} Padovani P., 2007, Ap\&SS, 309, 63
\bibitem[\protect\citeauthoryear{Padovani}{2007}]{b49} Padovani P., Giommi P. \& Rau A., 2012, MNRAS, 422, 48
\bibitem[\protect\citeauthoryear{Piner}{2008}]{b49} Piner B.G., Pant N. \& Edwards P.G., 2008, ApJ, 678, 64
\bibitem[\protect\citeauthoryear{Piner}{2010}]{b49} Piner B. G., Pant N. \& Edwards P.G., 2010, ApJ, 723, 1150
\bibitem[\protect\citeauthoryear{Piner}{2012}]{b49} Piner B.G., Pushkarev A.B., Kovalev Y.Y. et al., 2012, ApJ, 758, 84p
\bibitem[\protect\citeauthoryear{Piner}{2013}]{b49} Piner B.G. \& Edwards P.G., 2013, EPJ Web of Conferences, 6104021P
\bibitem[\protect\citeauthoryear{Plotkin}{2011}]{b49} Plotkin R.M., Markoff S., Trager S.C. \& Anderson S.F., 2011, MNRAS, 413, 805
\bibitem[\protect\citeauthoryear{Pushkarev}{2012}]{b49} Pushkarev A.B., Kovalev Y.Y., Lister M.L. \& Savolainen T., 2012, A\&A, 544, 3p
\bibitem[\protect\citeauthoryear{Rawlings \& Saunders}{1991}]{b50} Rawlings S. \& Saunders R., 1991, Nature, 349, 138
\bibitem[\protect\citeauthoryear{Sambruna et al.}{2010}]{b51} Sambruna R.M. et al., 2010, ApJ, 710, 24
\bibitem[\protect\citeauthoryear{Savolainen et al.}{2010}]{b51} Savolainen T. et al., 2010, A\&A, 512, A24
\bibitem[\protect\citeauthoryear{Sbarrato et al.}{2012}]{b52} Sbarrato T., Ghisellini G., Maraschi L. \& Colpi M., 2012, MNRAS, 421, 1764
\bibitem[\protect\citeauthoryear{Sbarrato et al.}{2014}]{b52} Sbarrato T., Padovani P. \& Ghisellini G., 2014, MNRAS, 445, 81
\bibitem[\protect\citeauthoryear{Scarpa \& Falomo}{1997}]{b53} Scarpa R. \& Falomo R., 1997, A\&A, 325, 109
\bibitem[\protect\citeauthoryear{Serjeant et al.}{1998}]{b54} Serjeant S., Rawlings S. \& Maddox S.J., 1998, MNRAS, 294, 494
\bibitem[\protect\citeauthoryear{Serjeant et al.}{1998}]{b54} Shen Y., Richards G.T., Strauss M.A. et al., 2011, ApJS, 194, 45
\bibitem[\protect\citeauthoryear{Taylor}{2007}]{b57} Taylor G.B. et al., 2007, ApJ, 671, 1355
\bibitem[\protect\citeauthoryear{Tchekhovskoy}{2011}]{b57} Tchekhovskoy A., Narayan R. \& McKinney J.C., 2011, MNRAS, 418, L79
\bibitem[\protect\citeauthoryear{Urry \& Padovani}{1995}]{b60} Urry C.M. \& Padovani P., 1995, PASP, 107, 803
\bibitem[\protect\citeauthoryear{Urry \& Padovani}{1995}]{b60} Veron-Cetty M.P. \& Veron P., 2010, A\&A, 518, 10
\bibitem[\protect\citeauthoryear{Wang et al.}{2004}]{b62} Wang J.M., Luo B. \& Ho L.C., 2004, ApJ, 615, L9
\bibitem[\protect\citeauthoryear{White}{2000}]{b62} White R.L., Becker R.H., Gregg M.D. et al., 2000, ApJS, 126, 133
\bibitem[\protect\citeauthoryear{Wills}{1995}]{b62} Wills B.J. \& Brotherton M.S., 1995, ApJ, 448, L81
\bibitem[\protect\citeauthoryear{Woo}{2005}]{b63} Woo J.H., Urry C.M., van der Marel R.P. et al., 2005, ApJ, 631, 762
\bibitem[\protect\citeauthoryear{Woo \& Urry}{2002}]{b63} Woo J.H. \& Urry C.M., 2002, ApJ, 579, 530
\bibitem[\protect\citeauthoryear{Wu}{2009}]{b63} Wu Z.Z., Gu M.F. \& Jiang D.R., 2009, RAA, 9, 168
\bibitem[\protect\citeauthoryear{Wu}{2014}]{b63} Wu Z.Z., Jiang D.R. \& Gu M.F., 2014, A\&A, 562, 64
\bibitem[\protect\citeauthoryear{Xie et al.}{2004}]{b67} Xie G.Z., Zhou S.B. \& Liang E.W., 2004, AJ, 127, 53
\bibitem[\protect\citeauthoryear{Xie et al.}{2007}]{b68} Xie G.Z., Dai H. \& Zhou S.B., 2007, AJ, 134, 1464
\bibitem[\protect\citeauthoryear{Xiong}{2014}]{b68} Xiong D.R. \& Zhang X., 2014, MNRAS, 441, 3375
\bibitem[\protect\citeauthoryear{Yee}{1978}]{b68} Yee H.K.C. \& Oke J.B., 1978, ApJ, 226, 753
\end{thebibliography}
\end{document}